\documentclass[%
 reprint,
superscriptaddress,
nofootinbib,
 amsmath,amssymb,
 aps,
 prd,
]{revtex4-2}

\newcommand\aap{A\&A}                
\newcommand\apjl{ApJ}                
\newcommand\jcap{J.~Cosmology Astropart. Phys.} 
\newcommand\mnras{MNRAS}             
\newcommand\physrep{Phys.~Rep.}      

\usepackage{graphicx}
\usepackage{dcolumn}
\usepackage{bm}

\usepackage{multirow}
\usepackage{booktabs}
\usepackage{orcidlink}

\newcommand{\IFAA}{Institute for Frontiers in Astronomy and Astrophysics, Beijing Normal University, Beijing 102206, China}
\newcommand{\SPA}{School of Physics and Astronomy, Beijing Normal University, Beijing 100875, China}
\newcommand{\IAP}{Institut d'Astrophysique de Paris, UMR~7095, CNRS, Sorbonne Universit\'e, 98~bis~boulevard Arago, 75014~Paris, France}
\newcommand{\CU}{Department of Physics, the Chinese University of Hong Kong, ST, NT, Hong Kong}
\newcommand{\SPLZL}{School of Physics and Laboratory of Zhongyuan Light, Zhengzhou University, Zhengzhou 450001, China}
\newcommand{\NAOC}{National Astronomical Observatories, Chinese Academy of Sciences, Beijing 100101, China}

\begin{document}

\preprint{APS/123-QED}

\title{Halo abundance and clustering in cosmologies with massive and asymmetric neutrinos}

\author{Yizhou Liu\,\orcidlink{0009-0005-8855-0748}}
\email{liuyz.cosmo@gmail.com}
\email{yzliu@bnu.edu.cn}
\affiliation{\IFAA}
\affiliation{\SPA}

\author{Wangzheng Zhang\,\orcidlink{0000-0003-0102-1543}}
\email{zhang@iap.fr}
\affiliation{\IAP}
\affiliation{\CU}

\author{Shihong Liao\,\orcidlink{0000-0001-7075-6098}}
\affiliation{\NAOC}

\author{Liang Gao\,\orcidlink{0009-0006-3885-9728}}
\affiliation{\IFAA}
\affiliation{\SPA}
\affiliation{\SPLZL}

\date{\today}

\begin{abstract}

Neutrinos are the most abundant fermions in the Universe and influence the formation of large-scale structure through both their non-zero masses and a possible chemical potential which can be described by a single asymmetry parameter. While most previous studies have focused on the impact of the neutrino mass, the role of neutrino asymmetry remains comparatively unexplored. In this work, we investigate how massive neutrinos ($M_{\nu}=0-0.24\,\mathrm{eV}$) with a non-zero asymmetry parameter ($\eta^{2}=0-0.8$) modify the halo mass function (HMF) and halo bias using cosmological N-body simulations with cosmological parameters consistently refitted to cosmic microwave background (CMB) observations. We find that at all redshifts, neutrino mass suppresses the abundance of massive halos, whereas neutrino asymmetry enhances the HMF over a broad mass range. Specifically, at $z=0$, the abundance of the most massive halos is reduced by up to $\sim30\%$ in the largest-mass case ($M_{\nu}=0.24\,\mathrm{eV}$), while neutrino asymmetry ($\eta^{2}=0.8$) produces a maximum $\sim5\%$ enhancement. These effects become increasingly pronounced at higher redshifts: by $z=4$ and $z=9$, the enhancement induced by neutrino asymmetry reaches $\sim25\%$ and $\sim75\%$, respectively, while the corresponding suppression due to neutrino mass deepens to below $\sim40\%$ and $\sim70\%$ of the massless case. For halo bias, we find that halos with masses above $10^{13.4}\,\mathrm{M_\odot}$ exhibit an enhanced large-scale bias due to neutrino mass, reaching up to $\sim5\%$ at $z=0$, while neutrino asymmetry reduces the bias by a few percent on linear scales. These trends strengthen with redshift, with the enhancement and suppression growing to $\sim15\%$ and $\sim10\%$ at $z=2$, respectively. Linear bias models provide an adequate, though not exact, description of halo bias in massive-neutrino cosmologies. Our results demonstrate that halo abundance and clustering offer sensitive probes of both neutrino mass and asymmetry.

\end{abstract}

\maketitle


\section{Introduction} \label{sec:introduction}

The large-scale structure (LSS) of the Universe provides a powerful laboratory for probing the fundamental properties of neutrinos \citep{Xia2012, Joudaki2013, Riemer2013, Zhao2013, PlanckCollaboration2014, Riemer2014}, yet such effects are often neglected in standard Lambda cold dark matter ($\Lambda$CDM) simulations. Relic neutrinos constitute the most abundant fermions in the Universe: they 
have large thermal velocities, thereby suppressing the growth of matter fluctuations below their free-streaming scale. As a consequence, neutrinos leave characteristic imprints on the statistical properties of cosmic structures across a wide range of scales and redshifts \citep{Brandbyge2010, Marulli2011, Costanzi2013, Castorina2014, Villaescusa-Navarro2014, Castorina2015, Inman2015, Villaescusa-Navarro2015, Adamek2017, Howlett2017, Liu2018, Dakin2019, Zeng2019, Bayer2021, Elbers2021, Massara2021, Wong2022, Zhou2022, Zhang2024, Sim2025}.

The impact of neutrinos on cosmic structure formation has been extensively investigated over the past decades. These studies have examined a variety of LSS observables, include the halo mass function (HMF) \citep{Brandbyge2010, Castorina2014, Castorina2015, Adamek2017, Liu2018, Bayer2021}, the halo bias \citep{Villaescusa-Navarro2014, Castorina2014, Castorina2015}, the matter power spectrum \citep{Dakin2019, Zeng2019, Elbers2021}, the marked power spectrum \citep{Massara2021}, the void size function \citep{Bayer2021, Verza2023}, the halo merger tree \citep{Liu2018, Wong2022}, the cosmic neutral hydrogen distribution \citep{Villaescusa-Navarro2015}, and matter velocity field \citep{Inman2015, Howlett2017, Zhou2022, Zhang2024}. Most of these works primarily focus on the dependence of LSS observables on the total neutrino mass, $M_\nu \equiv \sum_i m_i$, where the summation is over the three neutrino mass eigenstates.

Beyond the neutrino mass, an additional key parameter characterizing relic neutrinos is the chemical potential, which modifies their phase-space distribution as
\begin{eqnarray}
    f(p)=\frac{1}{e^{E(p)/T_\nu-\xi}+1},
\end{eqnarray}
where $E$ is the neutrino energy, and the dimensionless degeneracy parameter $\xi \equiv \mu/T_\nu$ is defined as the ratio of the chemical potential $\mu$ to the neutrino temperature $T_\nu$. The electron-type neutrino degeneracy is tightly constrained by Big Bang Nucleosynthesis (BBN), implying $\xi_e \ll 1$ \citep{Pitrou2018, Burns2023, Escudero2023}. In contrast, the degeneracy parameters of muon- and tau-flavor neutrinos can still reach $\mathcal{O}(1)$ according to current cosmic microwave background (CMB) constraints \citep{Yeung2021, Yeung2025}. Motivated by these considerations, and by the strong mixing between the $\nu_\mu$ and $\nu_\tau$ flavors, we set $\xi_e=0$ and assume $\xi_\mu=\xi_\tau$. Under these assumptions, the neutrino chemical potential can be described by a single asymmetry parameter, $\eta^2 \equiv \sum_i \xi_i^2$, where the sum runs over neutrino mass eigenstates. The relation between flavor and mass eigenstates is governed by the Pontecorvo-Maki-Nakagawa-Sakata (PMNS) matrix (see, e.g., \citealt{Barenboim2017}).

The degeneracy between the total neutrino mass $M_\nu$ and the asymmetry parameter $\eta^2$ has been explored using CMB observations \citep{Yeung2021, Yeung2025, Zhao2025}, the matter power spectrum \citep{Zeng2019}, halo merger trees \citep{Wong2022}, and pairwise velocity \citep{Zhang2024, Zhang2026Prep}. In this work, we focus on how massive neutrinos with non-zero asymmetry parameter modify the HMF and halo bias.
These statistics provide key probes of the abundance and clustering of collapsed structures and are widely used to test cosmological models and constrain neutrino properties \citep{Costanzi2013, Castorina2014, Villaescusa-Navarro2014}. Variations in neutrino properties generally require a refitting of cosmological parameters to maintain consistency with CMB observations \citep[e.g.][]{Zeng2019, Wong2022, Zhang2024}. In this framework, the effects of neutrino on LSS cannot be simply understood as the conventional small-scale power suppression that leads to a suppressed halo mass function and an enhanced halo bias \citep{Castorina2014}, thereby motivating a further detailed investigation of their impact on halo abundance and clustering.

We use the simulation suite presented in \citet{Zhang2024} to investigate the impact of neutrinos on the HMF and halo bias. We adopt the MICE fitting form, originally calibrated using the MICE simulation suite \citep{Crocce2010} and shown to provide an accurate description of halos at $z<1$ identified using the friends-of-friends (FoF) algorithm \citep{Castorina2014}, and refit it to extend its applicability to higher redshifts and to our adopted mass definition. We further examine how neutrino mass and asymmetry affect the HMF and halo bias across redshift. Finally, we test two classical bias models derived from peak-background split (PBS) theory \citep{Zentner2007, Baumann2022}, based on the Press-Schechter (PS) \citep{Press1974} and Sheth-Tormen (ST) \citep{Sheth1999, Sheth2002} mass functions.

This paper is organized as follows. Section~\ref{sec:Methodology} introduces the N-body simulations (\ref{sec:Simulation}) and the theoretical models for the HMFs (\ref{sec:hmf}) and halo bias (\ref{sec:hb}). Section~\ref{sec:Results} presents the impact of neutrinos on the HMF (\ref{sec:halo_abundance}) and halo bias (\ref{sec:halo_bias}) at different redshifts. Section~\ref{sec:conclusion} summarizes and discusses our findings.

\section{Methodology} \label{sec:Methodology}

\subsection{Simulation} \label{sec:Simulation}

\begin{table*}
\centering
\caption{Cosmological parameters of the $S_1$ and $S_2$ simulation suites. Here $M_\nu$ is the neutrino mass, and $\eta^{2}$ denotes the neutrino asymmetry. $H_{0}$ is the Hubble parameter. $\Omega_{c+b}$, $\Omega_{\Lambda}$, and $\Omega_{\nu}$ are the fractional energy densities of CDM plus baryons, dark energy, and neutrinos at redshift $z=0$, respectively. $n_{s}$ and $A_{s}$ are the spectral index and the amplitude of primordial scalar perturbations.}
\label{tab: cosmological parameter}
\begin{tabular}{c c c c c c c c}
\toprule
$M_\nu[\mathrm{eV}]$ & $\eta^{2}$ & $H_{0}[\mathrm{km\,s^{-1}Mpc^{-1}}]$ & $\Omega_{c+b}$ & $\Omega_{\nu}$ & $\Omega_{\Lambda}$ & $n_{s}$ & $A_{s}\,[10^{-9}]$ \\
\midrule
$0$ & $0$ & $68.03$ & $0.0307$ & $\sim 10^{-5}$ & 0.6930 & 0.9661 & 2.100 \\
\midrule
\multirow{3}{*}{0.06}
& 0   & $67.73$ & $0.3091$ & $0.0014$ & $0.6895$ & $0.9667$ & $2.102$ \\
& 0.4 & $68.77$ & $0.3061$ & $0.0014$ & $0.6925$ & $0.9728$ & $2.120$ \\
& 0.8 & $69.88$ & $0.3028$ & $0.0015$ & $0.6957$ & $0.9789$ & $2.138$ \\
\midrule
\multirow{3}{*}{0.15}
& 0   & $67.14$ & $0.3133$ & $0.0036$ & $0.6831$ & $0.9680$ & $2.106$ \\
& 0.4 & $68.18$ & $0.3102$ & $0.0037$ & $0.6861$ & $0.9741$ & $2.123$ \\
& 0.8 & $69.26$ & $0.3070$ & $0.0037$ & $0.6893$ & $0.9802$ & $2.143$ \\
\midrule
\multirow{3}{*}{0.24}
& 0   & $66.55$ & $0.3176$ & $0.0058$ & $0.6766$ & $0.9691$ & $2.107$ \\
& 0.4 & $67.58$ & $0.3143$ & $0.0059$ & $0.6798$ & $0.9752$ & $2.127$ \\
& 0.8 & $68.62$ & $0.3114$ & $0.0061$ & $0.6825$ & $0.9814$ & $2.146$ \\
\bottomrule
\end{tabular}
\end{table*}

Our dark-matter-only (DMO) simulations are based on a modified version of {\small GADGET-2} \citep{Springel2005}, extended to account for the effects of massive neutrinos, including the neutrino asymmetry \citep{Zeng2019}. Neutrino perturbations are treated using a linear response approach \citep{AliHaimoud2013}, in which the neutrino density contrast, $\delta_\nu$, is evolved by solving the Vlasov equation on a set of spatial grids and is coupled self-consistently to the nonlinear evolution of the cold dark matter (CDM) and baryon density contrast, $\delta_{c+b}$. The total matter overdensity entering the Poisson equation is therefore given by
\begin{equation}
    \delta_t = (1 - f_\nu)\,\delta_{c+b} + f_\nu\,\delta_\nu,
\end{equation}
where the neutrino energy density fraction is defined as $f_\nu \equiv \Omega_\nu / (\Omega_{c+b} + \Omega_\nu)$. $\Omega_{c+b}$ and $\Omega_{\nu}$ are the fractional energy densities of CDM plus baryons and neutrinos at redshift $z=0$, respectively.

Throughout this work, we assume three degenerate neutrino mass eigenvalues. A comprehensive description of the neutrino implementation in the N-body framework can be found in \citet{Zeng2019} and \citet{Zhang2024}. The initial conditions are generated using a modified version of {\small 2LPTic} \citep{Crocce2006}, in which the background expansion is modified to include neutrino contributions. Linear matter power spectra at the initial redshift $z=99$ are computed with {\small CAMB} \citep{Lewis2000}, incorporating neutrino asymmetry following \citet{Yeung2021}.

We consider a total of ten cases spanning a range of total neutrino mass $M_\nu$ and asymmetry parameter $\eta^2$. For each case, the remaining cosmological parameters are obtained by refitting the \emph{Planck} 2018 \texttt{plikHM\_TTTEEE} temperature and polarization likelihoods, together with the baryon acoustic oscillation (BAO) data set used in the \emph{Planck} 2018 analysis \citep{PlanckCollaboration2020}, using {\small CosmoMC}. The resulting mean values of cosmological parameters are summarized in Table~\ref{tab: cosmological parameter}.

Two sets of simulations are performed, denoted $S_1$ and $S_2$. Both suites employ $N_\mathrm{part} = 1024^3$ CDM\footnote{Below, the CDM includes both dark matter and baryon components.} particles and a particle-mesh grid with $N_{\rm grid} = 1024^3$, but differ in their box sizes, which are $L_{\rm box} = 1000\,h^{-1}\,\mathrm{Mpc}$ for $S_1$ and $250\,h^{-1}\,\mathrm{Mpc}$ for $S_2$. The corresponding particle masses are around $8\times10^{10}\,h^{-1}M_\odot$ and $1\times10^{9}\,h^{-1}M_\odot$, with gravitational softening lengths of $24.4\,h^{-1}\,\mathrm{kpc}$ and $6.1\,h^{-1}\,\mathrm{kpc}$, respectively. Here $h$ is defined through $H_0 = 100\,h\,\mathrm{km\,s^{-1}\,Mpc^{-1}}$.

Dark matter halos are identified using the {\small ROCKSTAR} halo finder \citep{Behroozi2013}. We restrict our analysis to distinct host halos, whose masses are defined using a spherical overdensity criterion with $\Delta = 200$ relative to the mean background CDM density. The halo center is calculated by the mean value of particles in the inner subgroup of this halo {\small ROCKSTAR}, and the expansion history adopted by {\small ROCKSTAR} is modified consistently to account for the presence of massive neutrinos. For the HMF and halo bias analysis, we only consider halos resolved with at least 200 particles, which sets the effective lower mass limit of our measurements.

\subsection{Halo mass function} \label{sec:hmf}

The HMF is commonly expressed in terms of the variance of the linear CDM density field, $\sigma(M,z)$, through the below function
\begin{equation}
    f(\sigma, z) = \frac{M}{\bar{\rho}}\frac{\mathrm{d}N(M,z)}{\mathrm{d}\mathrm{ln}\sigma^{-1}(M,z)},
	\label{eq:universal halo mass function}
\end{equation}
where $\bar{\rho}$ is the mean background CDM density and $N(M,z)$ denotes the comoving number density of halos more massive than $M$ at redshift $z$. 

The MICE fitting form for the HMF reads
\begin{equation}
    f_{\rm MICE}(\sigma, z) = A(z)\left[\sigma^{-a(z)}+b(z)\right]\exp\left[-\frac{c(z)}{\sigma^{2}}\right],
	\label{eq:MICE model}
\end{equation}
where the redshift-dependent parameters $A(z)$, $a(z)$, $b(z)$ and $c(z)$ are parameterized as power-law function of $(1+z)$ \citep{Crocce2010}. We refit these parameters using our neutrino cosmological simulation suite to obtain an accurate description of the halo abundance over the redshift range $0 \le z \le 9$ for our mass definition. 

The best-fitting parameters are obtained by minimizing a variance-weighted $\chi^{2}$ defined in logarithmic space\footnote{Throughout this paper, $\log$ denotes $\log_{10}$.},
\begin{equation}
    \chi^{2}(\theta) = \sum_{i=1}^{n} \left\{\frac{\log y_i - \log\left[\frac{\mathrm{d}N}{\mathrm{d}\log M}(M_{i},z_{i};\theta)\right]}{\sigma_{\log y_{i}}}\right\}^2,
	\label{eq:lsm}
\end{equation}
where $y_i$ and $\sigma_{\log y_i}$ denote the measured HMF and its associated Poisson uncertainty in logarithmic space, respectively, and $\theta$ represents the set of fitting parameters in Eq.~(\ref{eq:MICE model}).

\subsection{Halo bias} \label{sec:hb}

Dark matter halos are biased tracers of the underlying CDM density field. On sufficiently large scales, this relation can be described by a scale-independent linear bias factor $b$. In this work, we consider bias models based on the PBS framework, which assumes that the halo abundance can be expressed as a universal function of the peak height,
\begin{equation}
    \nu \equiv \frac{\delta_c}{\sigma(M,z)},
\end{equation}
where $\delta_c=1.686$ is the critical density for spherical collapse.

Under this assumption, the linear halo bias is given by
\begin{equation}
    b = 1 - \frac{1}{\delta_{c}}\frac{\mathrm{d}\mathrm{ln}f(\nu)}{\mathrm{d}\mathrm{ln}\nu},
	\label{eq:bias}
\end{equation}
where $f(\nu)$ denotes the universal function for different redshifts and cosmological models \citep{Zentner2007, Baumann2022}. For comparison, we adopt two widely used PBS-based bias models derived from the PS and ST mass functions \citep{Press1974, Sheth1999, Sheth2002}, respectively.

To facilitate a direct comparison with simulation measurements, we follow the procedure of \citet{Castorina2014} and compute the average bias of halos above a minimum mass threshold $M_{\mathrm{min}}$,
\begin{equation}
    b(>M_{\mathrm{min}}) = \frac{\int_{M_{\mathrm{min}}}b(M)n(M)\mathrm{d}M}{\int_{M_{\mathrm{min}}}n(M)\mathrm{d}M},
	\label{eq:mean bias}
\end{equation}
where $n(M)=\frac{\mathrm{d}N(M,z)}{\mathrm{d}M}$ is the comoving number density of halos
with mass $M$ at redshift $z$ corresponding to the chosen theoretical model.

For the PS and ST models, the above expression can equivalently be written in terms of $\nu$ as
\begin{equation}
    b(>\nu_{\mathrm{min}}) = \frac{\int_{\nu_{\mathrm{min}}}b(\nu)[f(\nu)/M(\nu)]\mathrm{d}\nu}{\int_{\nu_{\mathrm{min}}}[f(\nu)/M(\nu)]\mathrm{d}\nu},
	\label{eq:mean bias_nu}
\end{equation}
where the dependence on cosmology enters only through the mapping between $\nu$ and halo mass, $M(\nu)$. We compare these theoretical predictions with the measurement from simulations in Section~\ref{sec:Results}.

\section{Results} \label{sec:Results}

\begin{figure*}
\centering
\includegraphics[width=2\columnwidth]{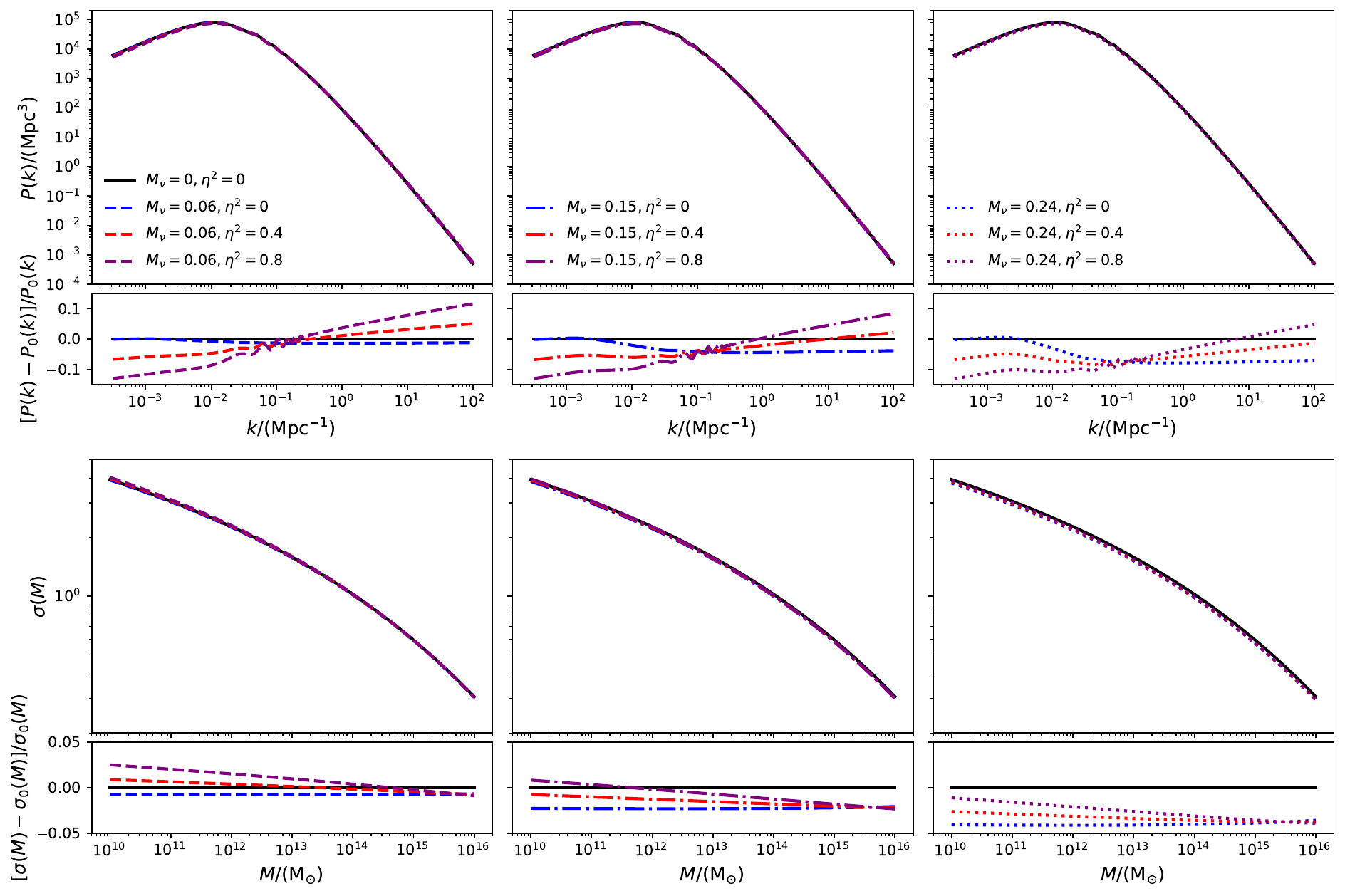}
\caption{Linear power spectrum (top panels) and mass variances (bottom panels) of CDM at $z=0$. Different colors and line styles correspond to different neutrino masses and asymmetry parameters, with the massless neutrino model adopted as the fiducial reference. Massive neutrinos suppress power spectra on small scales, while neutrino asymmetry parameter partially counteracts this suppression through the associated refitting of cosmological parameters required to maintain consistency with CMB observations.}
\label{fig:linear_pk}
\end{figure*}

\begin{figure*}
\centering
\includegraphics[width=2\columnwidth]{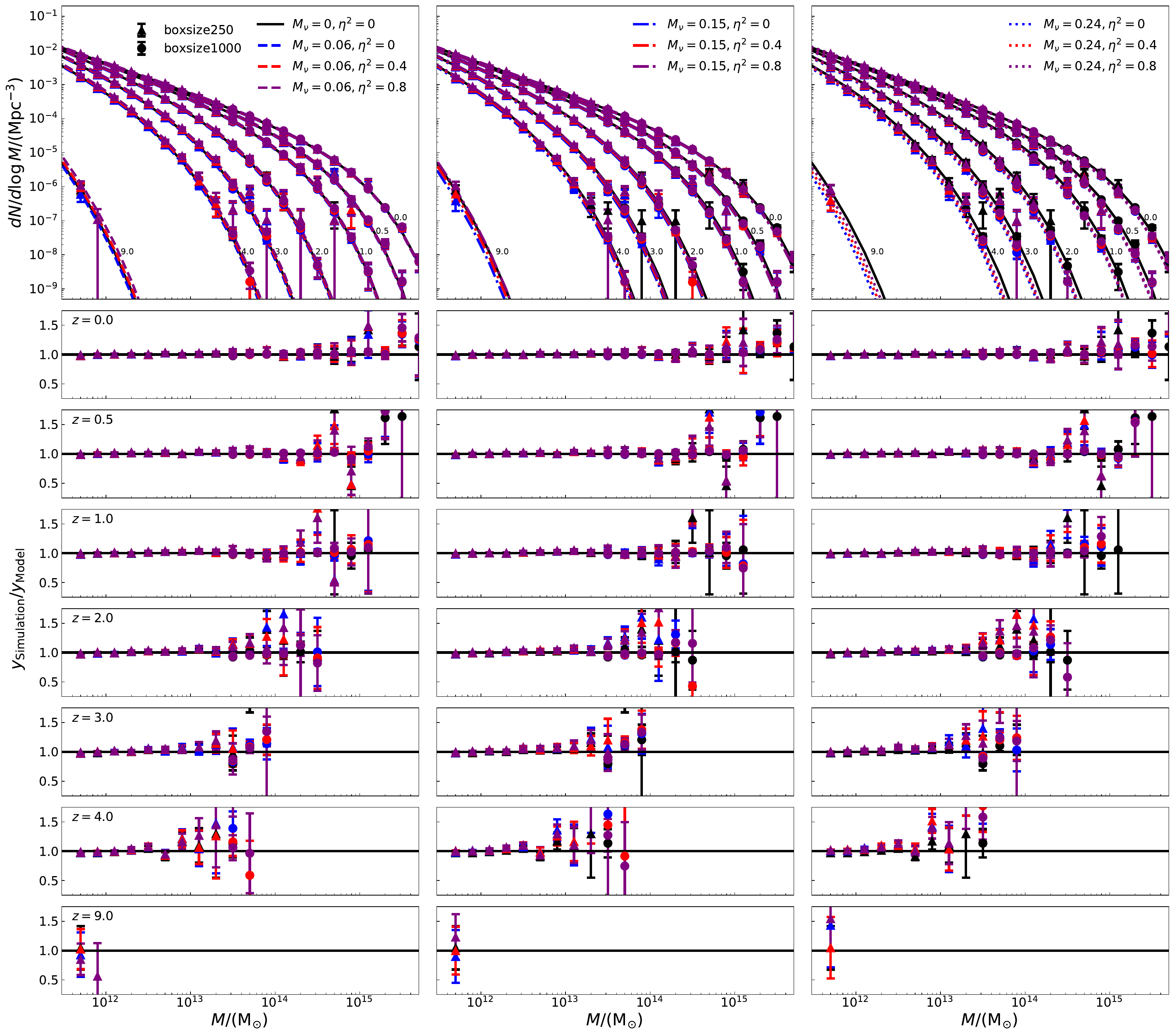}
\caption{HMFs across redshifts $z=0-9$. The top panels compare HMFs measured from the simulations with predictions from the refitted MICE model, while the lower seven rows of panels show the corresponding fractional residuals. Symbols denote simulation measurements and curves represent the refitted model predictions, with consistent colors indicating the same cosmology. Redshift values are labeled next to the corresponding curves in the top panels. The refitted MICE model accurately reproduces the HMF over the whole mass and redshift ranges probed, with deviations well within statistical uncertainties.}
\label{fig:hmf}
\end{figure*}

\begin{figure*}
\centering
\includegraphics[width=2\columnwidth]{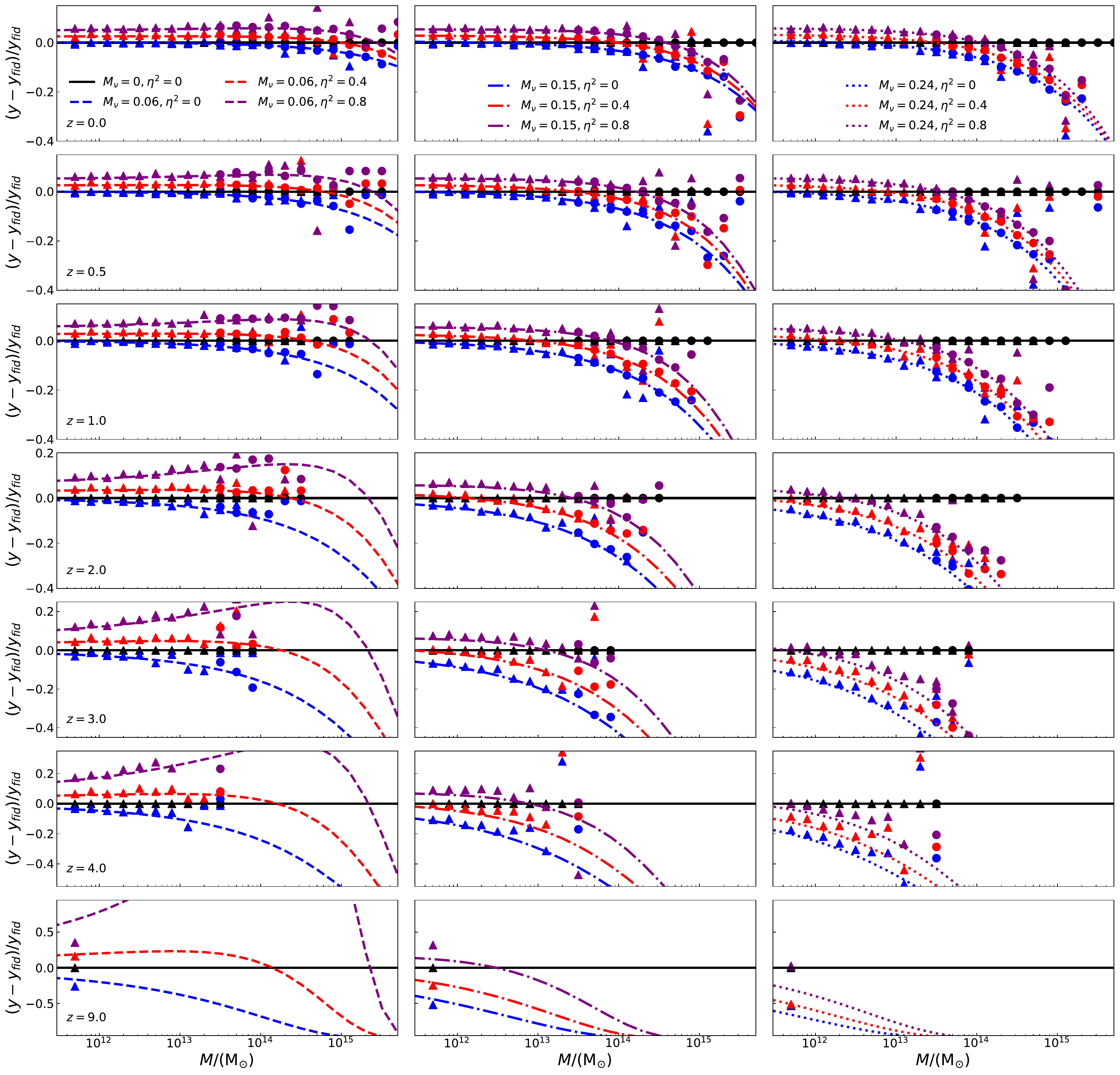}
\caption{Impact of neutrinos on the HMF. Shown are the relative effects of neutrino mass and asymmetry on the HMF, as predicted by the refitted model (curves) and measured from simulations (symbols), normalized to the massless neutrino case. The curves and symbols with the same color correspond to the same cosmology. Panels from top to bottom correspond to increasing redshift. Increasing the neutrino mass systematically suppresses the abundance of massive halos at all redshifts. In contrast, variations in the neutrino asymmetry act in the opposite direction, enhancing the halo abundance over a broad mass range.}
\label{fig:neutrino impact}
\end{figure*}

\begin{figure*}
\centering
\includegraphics[width=2\columnwidth]{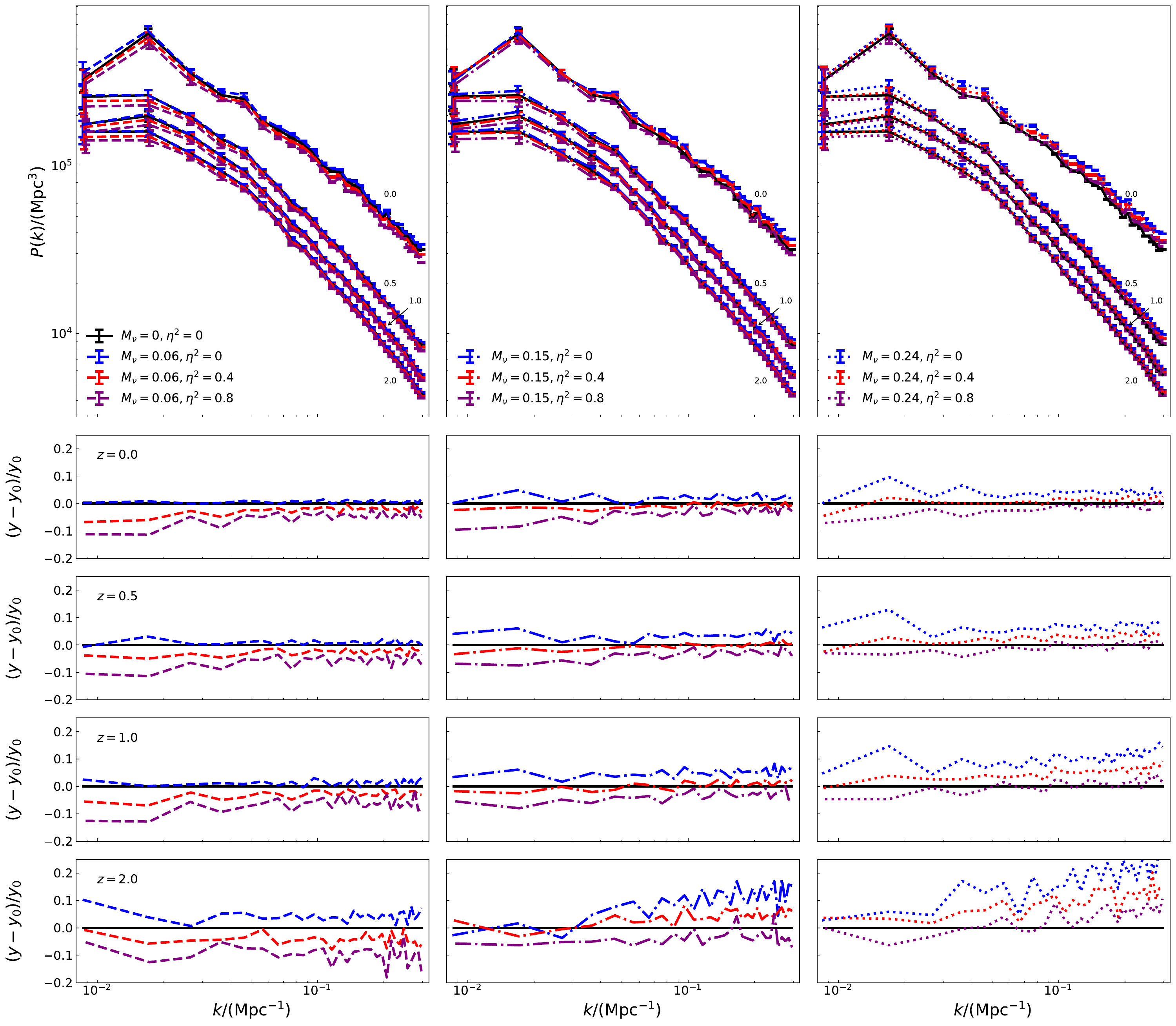}
\caption{Non-linear halo auto-power spectrum. The top panels show measurements from the $S_1$ simulation suite for halos with $M>10^{13.4}\,\mathrm{M_\odot}$ over redshifts $z=0-2$, with error bars indicating Poisson uncertainties and redshift labels placed next to each curve. The lower four rows of panels display the fractional differences relative to the massless neutrino case. Neutrino mass enhances the halo power spectrum amplitude, whereas a non-zero asymmetry parameter leads to a suppression, with model differences becoming increasingly pronounced toward higher redshift.}
\label{fig:halo pk}
\end{figure*}

\begin{figure*}
\centering
\includegraphics[width=2\columnwidth]{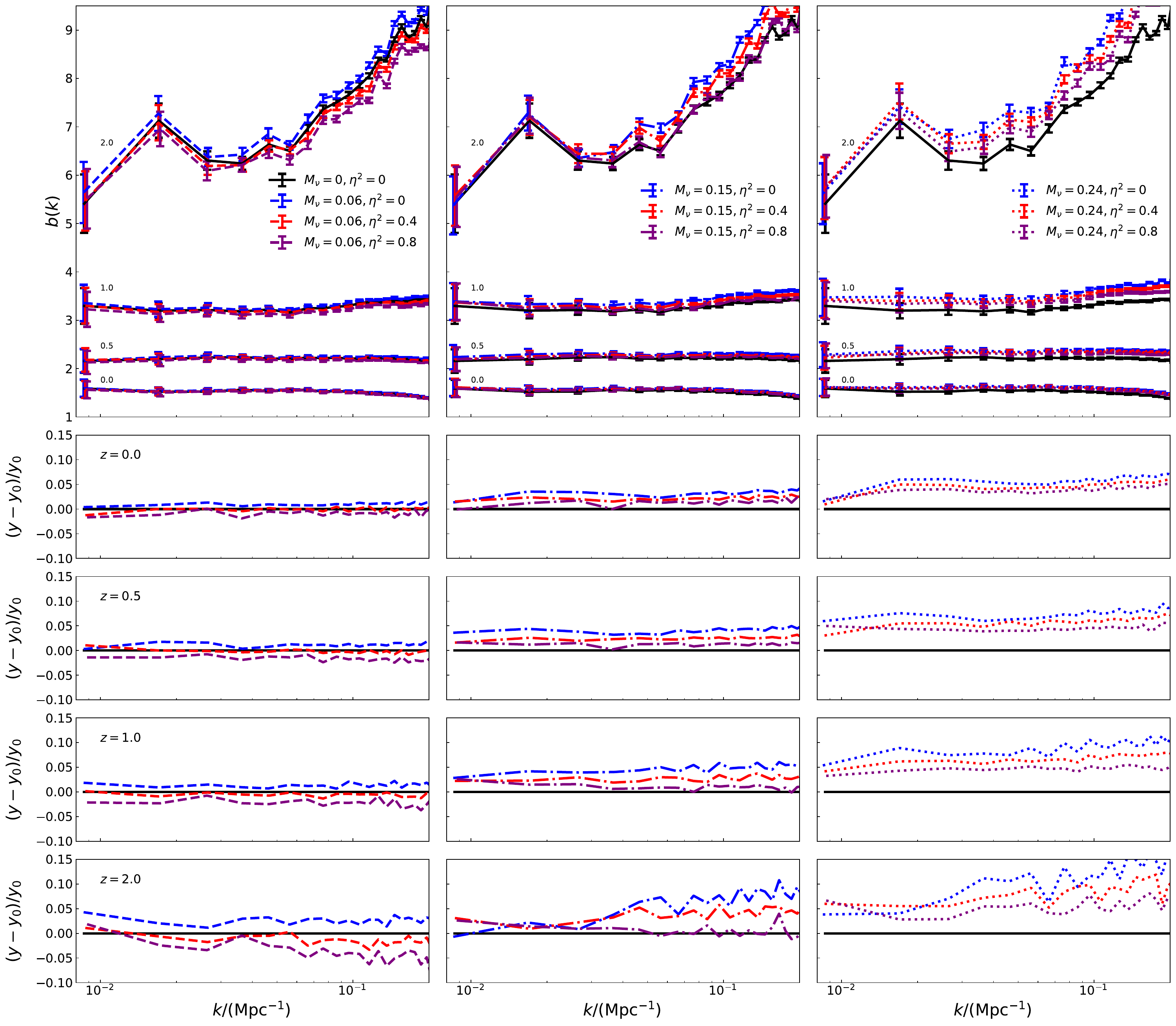}
\caption{Halo bias relative to the CDM density field. The top panels show the halo bias measured from the $S_1$ simulations for halos with $M>10^{13.4}\,\mathrm{M_\odot}$ over the redshift range $z=2$ to $z=0$, with the corresponding redshifts labeled in the upper left of each curve. The lower four rows of panels present fractional differences relative to the massless neutrino model. While the qualitative trends follow those seen in the halo power spectrum, the amplitudes differ, reflecting the combined influence of neutrinos on halo clustering and the underlying CDM density field.}
\label{fig:halo bias hp}
\end{figure*}

\begin{figure*}
\centering
\includegraphics[width=2\columnwidth]{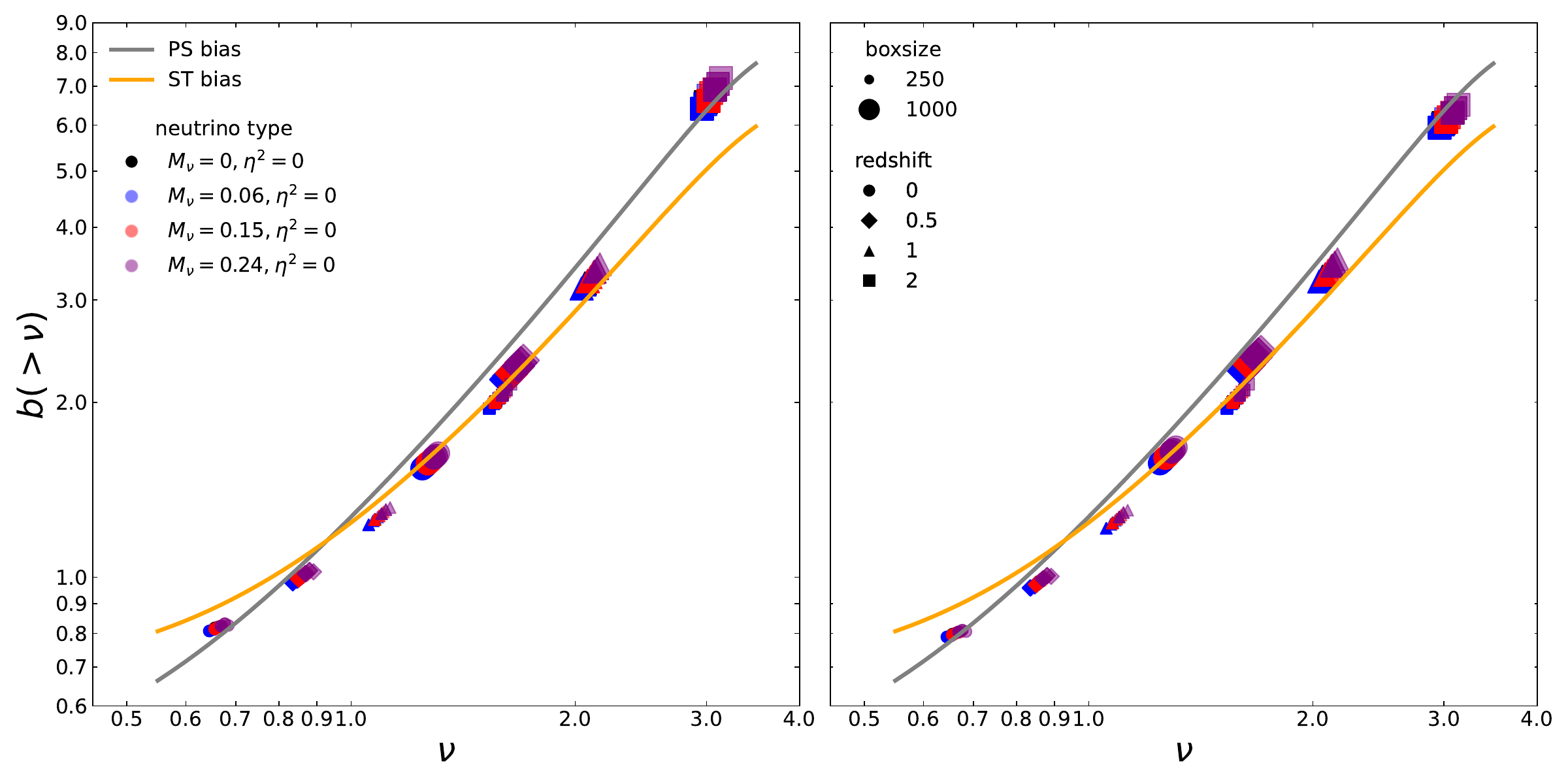}
\caption{Halo bias as a function of peak height $\nu$, averaged over the linear scale range $k=0.03-0.07\,\mathrm{Mpc^{-1}}$. Symbols with different markers, colors, and transparencies show halo bias measured from the $S_1$ and $S_2$ simulation suites. The left panel adopts $b=\sqrt{P_{\mathrm{hh}}/P_{\mathrm{cc}}}$, while the right panel uses $b=P_{\mathrm{hc}}/P_{\mathrm{cc}}$. Gray and orange curves correspond to the PS and ST predictions, respectively. Darker colors indicate larger neutrino chemical potentials. Linear bias models provide a reasonable, though not exact, description of halo bias in massive-neutrino cosmologies.}
\label{fig:bias nu}
\end{figure*}

\subsection{Halo abundance} \label{sec:halo_abundance}

In this section, we present the main results of our simulations. Since the abundance and clustering of dark matter halos are ultimately determined by the statistics of the underlying density field, we begin by examining the linear power spectrum and the corresponding mass variance, which provide the fundamental inputs for HMF modelling.

\subsubsection{Linear power spectrum \& mass variances}

The abundance of dark matter halos is primarily controlled by the variance of the linear density field smoothed on a mass scale $M$, $\sigma(M)$, which is in turn determined by the linear power spectrum. Therefore, before analysing the HMFs, it is instructive to quantify how different neutrino cosmologies modify these linear quantities.

Fig.~\ref{fig:linear_pk} shows the linear CDM power spectra (upper panels) and the corresponding mass variances $\sigma(M)$ (lower panels) for the different neutrino cosmological models considered in this work. All quantities are linearly extrapolated to redshift $z=0$. Different colors and line styles indicate different neutrino masses and asymmetry, while the massless neutrino simulation is adopted as the fiducial reference model. The lower sub-panels display the fractional differences relative to this fiducial case.

Massive neutrinos suppress power spectra on small scales due to their large thermal velocities, which inhibit clustering below the neutrino free-streaming scale. This suppression is clearly visible in the linear power spectrum at $k \gtrsim 0.01~{\rm Mpc}^{-1}$. This suppression becomes stronger for larger neutrino masses. It propagates directly into a reduction of $\sigma(M)$ for all mass halos shown here. Interestingly, the suppression of $\sigma(M)$ seems to be scale-independent for neutrino mass. However, at fixed neutrino mass, introducing a non-zero neutrino asymmetry parameter partially counteracts this suppression through the associated refitting of cosmological parameters required to maintain consistency with CMB observations, leading to an enhancement of the power spectrum at $k \gtrsim 0.1~{\rm Mpc}^{-1}$ and a systematic increase in $\sigma(M)$ for $M \lesssim 10^{15}\,\mathrm{M_{\odot}}$ relative to the zero-asymmetry-parameter case.

These trends have direct implications for the HMF. Since the halo abundance is exponentially sensitive to $\sigma(M)$, even percent-level modifications in the linear power spectrum can translate into significant changes in the number density of halos, particularly at the high-mass end. To analyze the impact of neutrinos on halo abundances, we first assess the performance of the theoretical HMF model by comparing it directly to simulation results.

\subsubsection{Refitting halo mass function}

Fig.~\ref{fig:hmf} presents the HMFs measured from the $S_1$ and $S_2$ simulation suites for halos with masses above $10^{11.6}\,\mathrm{M_\odot}$. Halos are grouped into logarithmic mass bins of width $0.2$ dex, and the number density in each bin is computed as $\Delta P/(V\,\Delta \log M)$, where $\Delta P$ is the number of halos within a given mass bin, $V$ is the comoving volume of the simulation box, and $\Delta \log M$ denotes the bin width in mass. Symbols with different colors and shapes represent the simulation measurements, with error bars indicating the statistical uncertainties estimated assuming Poisson noise. To describe these measurements, we adopt the MICE model and refit its parameters to our simulation results.

The resulting refitted model is given below, where the redshift evolution of the fitting parameters is described by simple power-law scalings,
\begin{equation}
\begin{aligned}
    A(z) &=  a_{1}(1 + z)^{a_{2}}\\
    a(z) &=  a_{3}(1 + z)^{a_{4}}\\
    b(z) &=  a_{5}(1 + z)^{a_{6}}\\
    c(z) &=  a_{7}(1 + z)^{a_{8}},
\end{aligned}
\label{eq:MICE parameters}
\end{equation}
where $a_{1}=0.6849$, $a_{2}=-0.3741$, $a_{3}=1.261$, $a_{4}=0.3612$, $a_{5}=0.2269$, $a_{6}=0.9253$, $a_{7}=1.1275$, and $a_{8}=0.05$. 

The best-fitting models are shown as curves in Fig.~\ref{fig:hmf}, while the lower panels display the fractional residuals between the refitted model and the simulation measurements at redshifts ranging from $z=0$ to $z=9$. Overall, the refitted MICE model provides an accurate description of the HMF across the full mass and redshift ranges probed by our simulations, with deviations remaining well within the statistical uncertainties. Having established the accuracy of the mass function model, we next focus on quantifying how neutrino mass and asymmetry parameter modify the halo abundance relative to the massless neutrino case across cosmic time.

\subsubsection{Impact of neutrino on HMF}

Fig.~\ref{fig:neutrino impact} compares the HMFs measured from simulations (symbols) with the predictions of the refitted MICE model (lines) for different neutrino masses and asymmetry parameters, over a wide range of redshifts. All results are shown relative to the massless neutrino case, allowing for a direct assessment of neutrino-induced modifications to halo abundances.

As shown in the figure, increasing the neutrino mass systematically suppresses the abundance of massive halos at all redshifts, which is consistent with previous works \citep{Castorina2014, Castorina2015, Adamek2017}. This suppression becomes more pronounced toward the high-mass end, reflecting the exponential sensitivity of the halo mass function to the variance of the density field. In contrast, variations in the neutrino asymmetry act in the opposite direction, enhancing the halo abundance over a broad mass range. The effects of neutrino mass and asymmetry are clearly non-degenerate. At fixed neutrino mass, increasing the asymmetry parameter leads to a nearly uniform upward shift of the mass function, whereas increasing the neutrino mass preferentially suppresses the high-mass tail. This qualitative difference persists across all redshifts probed by our simulations, indicating that these two neutrino properties imprint distinct signatures on halo abundances.

Quantitatively, at redshift $z=0$, a non-zero neutrino mass reduces the number density of the most massive halos ($\sim 2\times10^{15}\,\mathrm{M_{\odot}}$) by up to $\sim 30\%$ in the largest-mass case ($M_{\nu}=0.24\,\mathrm{eV}$) relative to the massless neutrino case. At the same time, introducing a maximum asymmetry parameter ($\eta^{2}=0.8$) increases the abundance of halos across the full mass range by approximately $5\%$. At higher redshifts, these effects become increasingly significant. By $z=4$, the asymmetry enhances the abundance of the most massive halos ($\sim 3\times10^{13}\,\mathrm{M_{\odot}}$) by up to $\sim 25\%$, while the corresponding neutrino mass suppresses their abundance to $>40\%$ of the massless case. At $z=9$, the enhancement due to the asymmetry reaches $\sim 75\%$ at the high-mass end ($\sim 5\times10^{11}\,\mathrm{M_{\odot}}$), whereas the suppression induced by neutrino mass deepens to $\sim 70\%$.

Overall, these results demonstrate that neutrino mass and asymmetry modify the HMF in distinct and redshift-dependent ways, with the strongest effects appearing at the halo massive end and at high redshift.

\subsection{Halo clustering} \label{sec:halo_bias}

While the HMF characterizes how neutrino properties affect halo abundances, a complementary and equally important aspect is how these properties modify the clustering of halos, which we quantify through the halo power spectrum and bias.

\subsubsection{Non-linear auto-power spectrum of halos} 

Fig.~\ref{fig:halo pk} shows the non-linear auto-power spectrum of halos with masses greater than $10^{13.4}\,{\mathrm{M_\odot}}$, measured from the $S_{1}$ simulation suite. Different colors and line styles correspond to different neutrino cosmological models. All results are shown relative to the massless neutrino case, which we adopt as the fiducial reference. The top panels present the halo power spectra, while the bottom four rows show the fractional residuals at redshifts $z=0$, $0.5$, $1.0$, and $2.0$. We focus on halos with masses above $10^{13.4}\,{\mathrm{M_\odot}}$ in $S_{1}$ simulation suite, since neutrino effects are known to be more pronounced for massive halos, whose clustering is more sensitive to changes in the growth of structure. Lower-mass halos in $S_{2}$ simulation suite exhibit qualitatively similar trends, but with a reduced amplitude of the neutrino-induced effects.

As shown in the figure, increasing the neutrino mass leads to an enhancement of the halo power spectrum on scale from $k=0.01\,\mathrm{Mpc^{-1}}$ to $k=0.3\,\mathrm{Mpc^{-1}}$. At fixed neutrino mass, introducing a finite neutrino asymmetry parameter produces the opposite effect, resulting in a suppression of the halo power spectrum amplitude. The differences between cosmological models become more pronounced as redshift increases. Especially, at the higher redshift $z=2$, the differences between models become more pronounced, with neutrino-induced deviations in the halo power spectrum growing from the $\sim 10\%$ level at $z=0$ to the $\sim 20\%$ level by $z=2$, depending on the neutrino mass and asymmetry, indicating that neutrino physics leaves an increasingly significant imprint on halo clustering at early times.

\subsubsection{Halo bias} 

To explore the neutrino-induced changes in halo bias, we next examine it in simulations. Fig.~\ref{fig:halo bias hp} presents the scale-dependent halo bias, defined as the ratio between the halo auto-power spectrum and the CDM auto-power spectrum,
\begin{equation}
b(k) = \sqrt{\frac{P_{\mathrm{hh}}(k)}{P_{\mathrm{cc}}(k)}},
\end{equation}
which quantifies how strongly dark matter halos trace the underlying density field. Here we still focus on halos with masses larger than $10^{13.4}\,\mathrm{M_{\odot}}$ from the $S_1$ simulation suite, where neutrino effects are more pronounced.

The top panels show the halo bias measured directly from the simulations at redshifts $z=0$ to $z=2$, while the lower panels display the relative deviation with respect to the massless neutrino model, which is still adopted as the fiducial reference. On scale from $k=0.01\,\mathrm{Mpc^{-1}}$ to $k=0.2\,\mathrm{Mpc^{-1}}$, the relative deviation exhibits only weak scale dependence, indicating that the neutrino-induced modifications predominantly affect the overall amplitude rather than introducing strong scale-dependent features in the linear regime. As expected, the impact of neutrinos on halo bias becomes increasingly significant at higher redshifts. For a fixed asymmetry parameter, increasing the neutrino mass systematically enhances the halo bias, whereas at fixed neutrino mass, increasing the asymmetry parameter leads to a reduction of the bias amplitude. This trend is already visible at $z=0$ and becomes progressively stronger toward $z=2$, consistent with the redshift evolution observed in the halo power spectrum.

However, although the qualitative trends of the halo bias are consistent with those observed in the halo power spectrum, their quantitative amplitudes differ, reflecting the combined impact of neutrinos on both halo clustering and the underlying power spectrum. Since the halo bias is defined through the ratio between halo and CDM power spectra (see Appendix), its response to neutrino properties cannot be directly inferred from halo clustering alone. Quantitatively, we find that increasing the neutrino mass leads to an enhancement of the halo bias up to the level of $\sim 5\%$ on these scales at $z=0$, with the effect growing to $15\%$ by $z=2$, depending on the neutrino mass. In contrast, increasing the neutrino asymmetry parameter systematically reduces the halo bias, with a suppression of a few percent at low redshift and reaching $\sim 10\%$ at $z=2$. For massive neutrinos with zero asymmetry parameter, a similar trend in the halo bias can be seen in the results of \citet{Castorina2015}.

We extract the halo bias averaged over the scale range $k=0.03-0.07$ Mpc$^{-1}$ and study its dependence on the peak height $\nu$, comparing the simulation results with the linear-theory predictions from the PS and ST bias models. As shown in Fig.~\ref{fig:bias nu}, the solid lines represent the theoretical predictions, while the symbols correspond to measurements from the $S_1$ and $S_2$ simulation. The left panel shows the bias defined as $b=\sqrt{P_{\mathrm{hh}}/P_{\mathrm{cc}}}$, whereas the right panel adopts the alternative definition $b=P_{\mathrm{hc}}/P_{\mathrm{cc}}$. These two estimators differ slightly due to stochasticity and shot-noise contributions, but are expected to coincide on sufficiently large, linear scales.

We find that both PS and ST models reproduce the overall trend of the measured bias as a function of $\nu$, although noticeable deviations remain. The level of agreement is nevertheless reasonable given the simplicity of the analytic models. While Eq.~\ref{eq:mean bias_nu} shows that the average bias exhibit a mild dependence on cosmology, here we only display the theoretical prediction computed in the massless-neutrino cosmology for clarity. The weak cosmological dependence implies that this choice does not affect our qualitative conclusions. Overall, this comparison indicates that linear bias models provide an adequate, though not exact, description of halo bias in massive-neutrino cosmologies.

\section{Conclusion and Discussion} \label{sec:conclusion}

In this work, we have investigated the impact of neutrino properties on halo statistics using a suite of cosmological simulations in which the background cosmological parameters are consistently refitted to CMB data for each neutrino model. By varying both the total neutrino mass $M_{\nu}$ ($0-0.24\,\mathrm{eV}$) and the neutrino asymmetry parameter $\eta^{2}$ ($0-0.8$), we analysed how neutrino physics affects the HMF and halo bias across a wide range of redshifts. This approach allows us to analyze the combined effects of neutrinos on structure formation while maintaining consistency with CMB constraints, and provides a controlled framework to assess how neutrino-induced modifications of the CDM density field propagate into halo abundance and clustering.

We first examined the impact of neutrinos on the HMF. Increasing the neutrino mass leads to a systematic suppression of halo abundance, with the effect being strongest for massive halos. In contrast, a non-zero neutrino asymmetry enhances the halo abundance over a broad mass range, partially compensating for or even outweighing the suppression induced by neutrino mass. These trends reflect the combined influence of neutrino free-streaming, refitted cosmological parameters and modified expansion history. As redshift increases, the relative differences in halo abundance between different neutrino models grow: at $z=0$, the abundance of the most massive halos is reduced by up to $\sim30\%$ in the largest-mass case ($M_{\nu}=0.24\,\mathrm{eV}$), while introducing a maximum asymmetry parameter ($\eta^{2}=0.8$) would produce a $\sim5\%$ enhancement. By $z=4$ and $z=9$, the enhancement induced by neutrino asymmetry reaches $\sim25\%$ and $\sim75\%$, respectively, while the corresponding suppression due to neutrino mass deepens to below $\sim40\%$ and $\sim70\%$ of the massless case. These results indicate that neutrino effects on halo formation are amplified at earlier cosmic times.

We then studied the impact of neutrino properties on halo bias, focusing on massive halos for which neutrino effects are most evident. We find that massive neutrinos tend to increase the halo bias, while a larger neutrino asymmetry generally reduces it. The magnitude of these effects increases with redshift, whereas the scale dependence remains weak on large linear scales. At $z=0$, halos with masses above $10^{13.4}\,\mathrm{M_\odot}$ exhibit an enhanced large-scale bias due to neutrino mass, reaching up to $\sim5\%$ at $z=0$, while neutrino asymmetry reduces the bias by a few percent. By $z=2$, the enhancement and suppression grow to $\sim15\%$ and $\sim10\%$, respectively. Linear bias models provide an adequate, though not exact, description of halo bias in massive-neutrino cosmologies.

Taken together, our results highlight a coherent physical picture in which neutrinos first modify the CDM power spectrum, thereby altering the formation and abundance of dark matter halos, and ultimately reshaping halo bias and clustering. The distinct and non-degenerate impacts of neutrino mass and neutrino asymmetry on halo statistics suggest that measurements of halo abundance and large-scale clustering would provide avenues to constrain neutrino properties. Future observational probes of LSS, particularly those sensitive to massive halo abundance and their clustering at intermediate to high redshifts, could therefore offer valuable constraints on neutrino physics beyond the standard cosmological model.

\begin{acknowledgments}

We thank Ming-chung Chu for helpful discussions and valuable comments on this work. We would like to thank Hang Yang, Tianyu Zhang, Haonan Zheng, and Rui Hu for their help and discussions. We acknowledge the supports from the National Natural Science Foundation of China (Grant No. 12588202) and the National Key Research and Development Program of China (Grant No. 2023YFB3002500). WZ acknowledges the support of the French Agence Nationale de la Recherche (ANR), under grant ANR-23-CE31-0010 (project ProGraceRay). The computational resources used for the simulations in this paper were provided by the Central Research Computing Cluster at the Chinese University of Hong Kong. This research is supported by grants from the Research Grants Council of the Hong Kong Special Administrative Region, China, under Project Nos.\ AoE/P-404/18 and 14300223.

\end{acknowledgments}

\nocite{*}

\bibliography{apssamp}

\appendix

\section{Non-linear auto-power spectrum of CDM}

\begin{figure*}
\centering
\includegraphics[width=2\columnwidth]{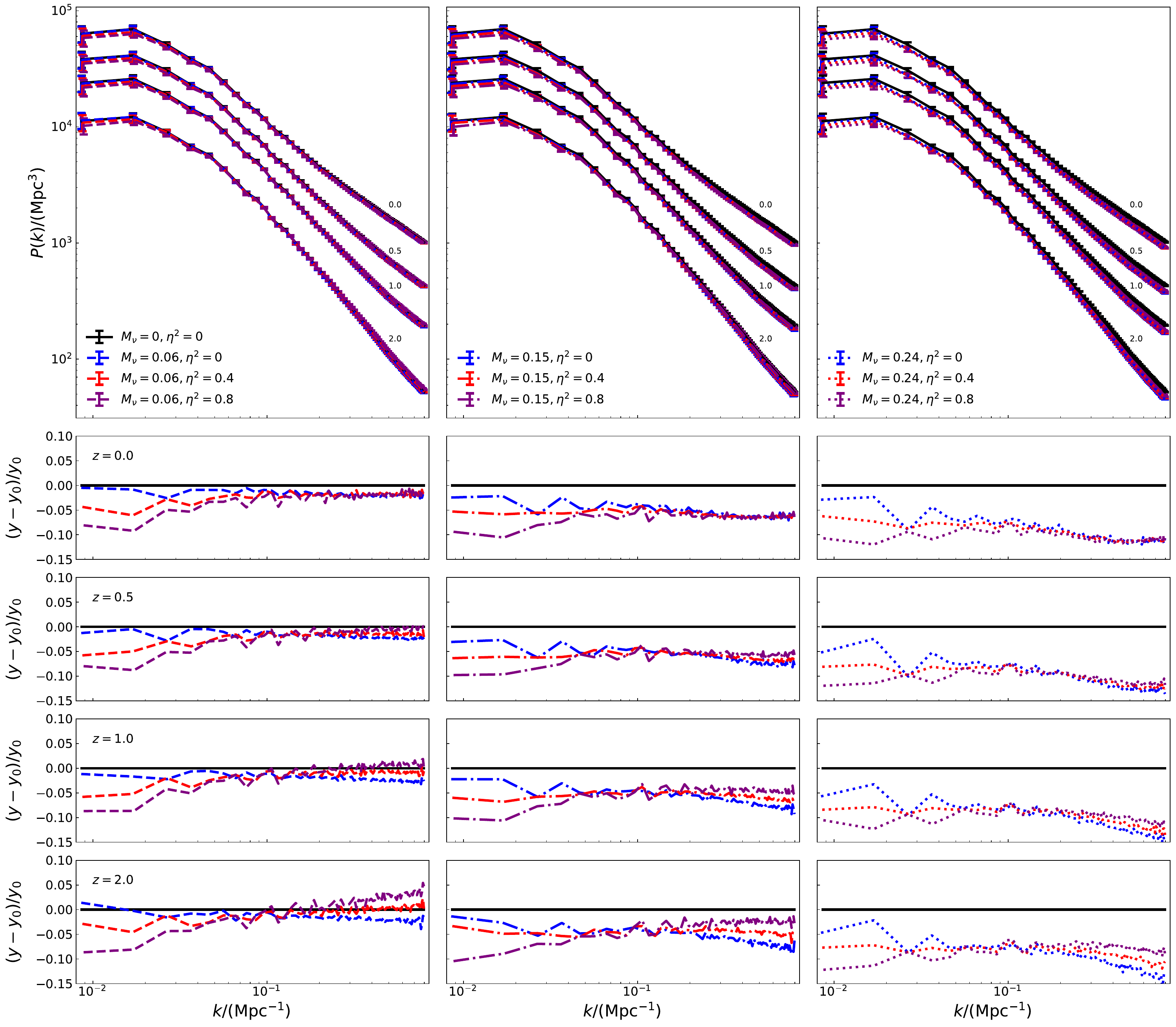}
\caption{Non-linear CDM auto-power spectrum. The top panels show measurements from the $S_{1}$ simulation suite for CDM over redshifts $z=0-2$, with redshift labels indicated next to each curve. The lower four rows of panels display fractional differences relative to the massless neutrino case. A non-zero neutrino mass leads to a suppression of the power spectrum over the entire range of scales considered, while the presence of a neutrino asymmetry produces a scale-dependent modification: the power spectrum is suppressed on large scales ($k \lesssim 0.1\,\mathrm{Mpc^{-1}}$) and enhanced on smaller scales ($k \gtrsim 0.1\,\mathrm{Mpc^{-1}}$), particularly at high redshift. By $z=0$, the residual curves corresponding to different asymmetry parameters converge for $k \gtrsim 0.1\,\mathrm{Mpc^{-1}}$, while remaining nearly unchanged on larger scales.}
\label{fig:particle pk}
\end{figure*}

Fig.~\ref{fig:particle pk} presents the non-linear auto-power spectrum of CDM measured from the $S_{1}$ simulation suite over the redshift range $z=0-2$. Different colors and line styles correspond to different neutrino cosmological models. All results are shown relative to the massless neutrino case, which is adopted as the fiducial reference. The top panels display the absolute power spectra, while the lower four rows show the fractional residuals at redshifts $z=0$, $0.5$, $1.0$, and $2.0$. 

A non-zero neutrino mass leads to a suppression of the power spectrum over the entire range of scales considered, with the effect becoming more pronounced toward higher redshift. The presence of a neutrino asymmetry produces a scale-dependent modification: the power spectrum is suppressed on large scales ($k \lesssim 0.1\,\mathrm{Mpc^{-1}}$) and enhanced on smaller scales ($k \gtrsim 0.1\,\mathrm{Mpc^{-1}}$), particularly at high redshift. At $z=2$, the residuals of the non-linear power spectrum closely track those of the linear power spectrum shown in Fig.~\ref{fig:linear_pk}. As the Universe evolves, non-linear gravitational interactions progressively modify these residuals. By $z=0$, the residual curves corresponding to different asymmetry parameters converge for $k \gtrsim 0.1\,\mathrm{Mpc^{-1}}$, while remaining nearly unchanged on larger scales. Together with the corresponding halo power spectra, the CDM power spectrum determines the scale- and redshift-dependent behaviour of the halo bias discussed in the main text.

\end{document}